# High-Speed Time Series Prediction with a GHz-rate Photonic Spiking Neural Network built with a single VCSEL

Dafydd Owen-Newns[1], Lina Jaurigue[2], Joshua Robertson[1], Andrew Adair[1], Jonnel Anthony Jaurigue[2], Kathy Lüdge[2], and Antonio Hurtado[1]

[1]Institute of Photonics, SUPA Department of Physics, University of Strathclyde, 99 George Street, G1 1RD, Glasgow, UK.
[2]Technische Universität Ilmenau, Institut für Physik, Weimarer Straße 25, 98693 Ilmenau, Germany
*Corresponding author. Email: dafydd.owen-newns@strath.ac.uk

**Abstract**

Photonic technologies hold significant potential for creating innovative, high-speed, efficient and hardware-friendly neuromorphic computing platforms. Neuromorphic photonic methods leveraging ubiquitous, technologically mature and cost-effective Vertical-Cavity Surface Emitting Lasers (VCSELs) are of notable interest. VCSELs have demonstrated the capability to replicate neuronal optical spiking responses at ultrafast rates. These characteristics have triggered research into applying these key-enabling devices in spike-based photonic computing. Here, a GHz-rate photonic Spiking Neural Network (p-SNN) using a single VCSEL is reported, and its application to a complex time-series prediction task is demonstrated for the first time. The VCSEL p-SNN combined with a technique to induce network memory, is applied to perform multi-step-ahead predictions of a chaotic time-series. By providing the feedforward p-SNN with only two temporally separated inputs excellent accuracy is experimentally demonstrated over a range of prediction horizons. VCSEL-based p-SNNs therefore offer ultrafast, efficient operation in complex predictive tasks whilst enabling hardware implementations. The inherent attributes and performance of VCSEL p-SNNs hold great promise for use in future light-enabled neuromorphic computing hardware.

## Introduction

Artificial Neural Networks (ANNs) form a core component of many implementations of Artificial Intelligence (AI) due to their demonstrated capability to deliver high-level performance in various complex information processing tasks like pattern recognition, data classification, and image and language processing. ANNs are inspired by the functioning of biological neurons in the brain, consisting of parallel arrays of large numbers of nodes (neurons) that use non-linear transformations to achieve efficient processing and decision-making. Neuromorphic (brain-like) processing aims at reproducing and exploiting the spike-based signalling in biological neurons for computation and has gathered extensive research interest in recent years. Driven by the desire to find novel computing architectures not limited by the challenges of traditional digital computing platforms, research into neuromorphic processing has led to spike-operating electronic processing systems such as Loihi 2 [1], TrueNorth [2], Spinnaker 2 [3] and BrainScaleS-2 [4] (see [5] for a review).

Photonic systems, offer further avenues of development in neuromorphic computing [6] that make use of the fundamental advantages intrinsic to the optical medium (e.g. high bandwidth, reduced cross-talk, long communication links, etc.) and that can help overcome inherent limitations arising in electronic-based platforms. Optical devices offer high data throughput due to the capability to transmit many signals in the same channel without interference, granting high computation densities and parallelism via wavelength and mode multiplexing techniques. Additionally, photonics has access to a range of linear and nonlinear interactions that can be implemented with high efficiency and leveraged for information processing functionalities. Furthermore, photonics offers prospects for the implementation of low power data links, with faster baseline operating speeds, for a promising highly energy efficient neuromorphic computing platform.

The realisation of such neural networks using photonic technologies have been shown recently using a number of systems such as optical modulators [7], micro-ring weight banks [8], phase change materials [9] and semiconductor lasers [10, 11]. Semiconductor lasers are of particular interest due to their dynamical behaviour, which allow for the implementation of neuronal-like responses at ultrafast speeds. In particular, one type of semiconductor laser, the Vertical-Cavity Surface-Emitting Laser (VCSEL) has been the subject of extensive research as it has been demonstrated to exhibit spiking dynamics similar to those of biological neurons but at ultrafast sub-nanosecond speeds.

VCSELs are industrially ubiquitous semiconductor lasers, found in a wide variety of consumer devices (e.g. mobile phones, automotive sensors), as well as in telecommunications and data centres. Interestingly, VCSELs operating in the 1310 and 1550 nm telecom windows have demonstrated excitable neuronal behaviours, emitting controllable optical spikes in response to optical signals injected into the laser cavity [12]. The neural-like optical spikes produced by VCSELs are short (~ 100ps-long) and can be triggered at rates up to a few GHz. They also exhibit other useful neuromorphic properties such as spike-firing thresholding, integration-and-fire responses and refractoriness [13–15]. These properties have been used for several purposes: implementing logical XOR operation [16]; and image processing tasks ranging from pattern recognition [17–19] to rate-coding [20]. Due to the maturity of VCSEL technology, this increasingly promising approach to photonic artificial spiking neurons can be realised with off-the-shelf optical components at key telecom wavelengths and in hardware-friendly configurations, helping remove the barrier of expensive bespoke designs.

One established use of VCSELs for ANNs is in photonic-based reservoir computing (RC). These are large scale, recurrent neural networks in which all internal connections and input layer weights are fixed, and only the output layer weights require training. Photonic RC systems have demonstrated good performance in several complex tasks, and as the parameters and connectivity of the reservoir are fixed, they are ideal to implement in purpose-built photonic hardware. Photonic RC systems have been implemented in recent years using optoelectronic and a wide range of optical devices [21-26], including micro-ring resonators, photonic integrated systems and semiconductor lasers. In particular, VCSEL-based photonic RC systems, have attracted important research attention due to their dual linear polarisation emission and lower energy operation, which can be leveraged to improve computing performance [27, 28]. VCSEL photonic RC systems have been constructed using space- [29, 30], and time-multiplexing [31, 32] paradigms, in which the nodes and connections of the network are defined by the time varying (or spatial) dynamics of light in the laser cavity (see [33] for a review of VCSEL-based approaches to RC).

Recently, the optical spiking dynamics exhibited by VCSELs have been utilised in combination with time-multiplexing techniques to construct a photonic Spiking Neural Network (p-SNN) [34]. This p-SNN was built with a single time-multiplexed VCSEL, the network connections and the nonlinear transformation were provided by the excitable optical spiking dynamics of the device. These link information between many network nodes for effective data processing, demonstrating excellent performance on classification tasks, as they enable separating data with complex decision boundaries [34]. Importantly, the use of optical spiking signals to process information, offers key benefits when compared to traditional non-spiking photonic neural network implementations. These include the possibility to use novel training algorithms that only require the training of very few network nodes (increasing training simplicity and overall processing speed), they allow operation with sparse signalling and weight matrices, and they can perform accurately even when low dataset sizes are used for training [35].

However, to date these p-SNNs lack internal memory and cannot therefore operate in key tasks, such as time-series prediction, which are of strategic importance in a variety of sectors (e.g. predictive maintenance, weather forecasting, etc.). This work tackles this fundamental challenge by demonstrating for the first time, the successful operation of a p-SNN built with a single, commercially-sourced, telecom-wavelength VCSEL on a chaotic timeseries prediction task. Specifically, we demonstrate experimentally the p-SNN's successful execution of multi-step-ahead prediction of the chaotic Mackey-Glass time series. This task requires that the system produces an output corresponding to a real number (the time-series value), and that the system has knowledge of previous temporal data inputs (memory). This requires going beyond the simpler operation in classification tasks for which memory is not required and whose output is limited to an element of a finite set (equal to the number of datapoint classes). Here, we demonstrate the successful performance of the p-SNN overcoming these fundamental challenges caused by the representation of real numbers with discrete neural-like spikes, and the lack of (memory enhancing) feedback network connections. We achieve this by applying a memory-inducing technique in the pre-processing of the input data that has recently been shown to be an effective means of improving the performance of reservoir computers on tasks requiring memory [36,43].

## Photonic Spiking Neural Network (p-SNN)

The p-SNN created in this work deploys concepts from spiking and time-multiplexed neural networks as well as from the Reservoir Computing (RC) paradigm. Unlike standard neural network approaches that implement in hardware many (spatial) nodes and network connections, time-multiplexed systems require only one non-linear (spatial) node for operation. Time-multiplexing involves treating the time-varying output intensity of a single non-linear node as the output of many distinct non-linear (virtual) neurons. By sampling the non-linear device at a discrete temporal separation (referred to as θ) the output for a set number of virtual nodes ($N_v$) can be interpreted, realising a virtual neural network (see [37] for a review on time-multiplexed photonic RC). In this work a single telecom 1300nm-VCSEL yielding a non-linear temporal high-speed spiking optical output is used in this way to realise an entire p-SNN of hundreds of (virtual) spiking neurons. Figure 1(a) provides the experimental setup developed to implement and investigate the VCSEL-based p-SNN of this work (Methods section). Figure 1(b) plots a schematic diagram of the node coupling scheme and architecture of the p-SNN for its use in the chaotic time-series prediction tasks.

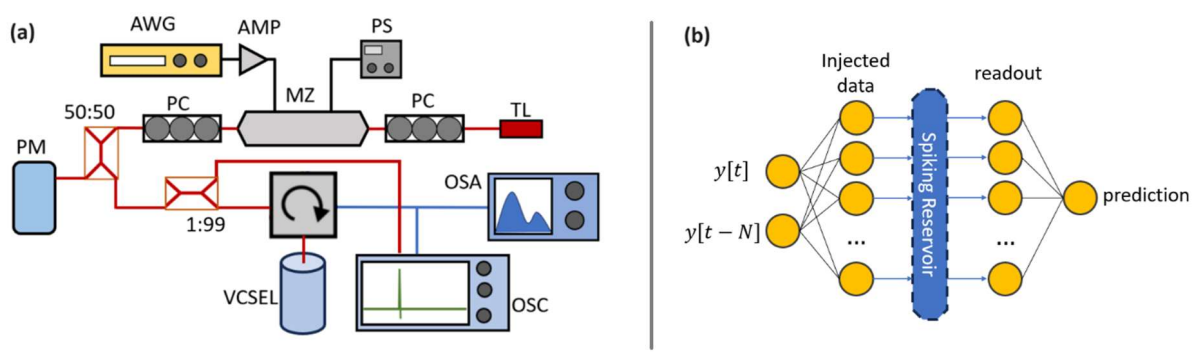

Figure 1: (a) Diagram of the experimental photonic spiking neural network. A tuneable laser (TL) provides light which is intensity modulated by a Mach Zehnder (MZ) modulator. The input to the MZ modulator is given by a DC bias (generated by a Power Source, PS) and an arbitrary wave generator (AWG) followed by an RF amplifier (AMP), encoding the data to be processed into the time varying intensity of the light. Polarisation controllers (PC) are used to match the light polarisation of the input signal to that of the MZ modulator and the VCSEL. The input signal is monitored with a Power Meter (PM) and oscilloscope (OSC). The modulated light signal is then injected into the VCSEL where it can trigger ultrafast optical spikes. The VCSEL's output is read by an optical spectrum analyser (OSA) and a photodiode and oscilloscope (OSC). (b) Schematic diagram of the architecture of the VCSEL-based p-SNN [34-35].

In our experiments, the selected individual node time was equal to θ = 250 ps (approximately one quarter of the 1ns refractory period of this VCSEL's spiking behaviour); thus, readily allowing without any further optimisation to operate at 4 GHz input data rates. Additionally, we investigated experimentally three different p-SNN sizes with total node counts of $N_v$ = 256, 512 and 1024. In the p-SNN, the connections between (virtual) nodes in the network are fixed and are dictated by the temporal carrier dynamics of the spiking VCSEL [34]. Therefore, in this type of p-SNN, as in the RC paradigm, both the input and hidden layer connections are fixed and only the output layer weights required training. This allows for highly-reduced training protocols compared with fully trained networks, whilst simultaneously permitting hardware-friendly implementations and high-accuracy in key processing tasks (for a review on VCSEL-based photonic RC systems see for example [33] and references therein).

Recent works [34-35] have reported the application of VCSEL-based p-SNNs to perform complex classification tasks, achieving very good performance [34-35]. Notably, the use of ultrafast optical spikes in the p-SNN to process information, offered important additional advantages, including the ability to operate with very small training dataset sizes (down to just a few examples), the ability to tune its node count and change therefore the network architecture, and the possibility to operate with highly-efficient learning protocols requiring only the use of less than 1% of the output layer nodes for operation [34-35]. However, to date these p-SNNs lack internal system memory, and therefore their use in strategic memory-requiring processing functionalities has remained elusive. The present work tackles this crucial challenge demonstrating for the first time an induced memory technique in the p-SNN of this work to enable its successful operation in memory-requiring chaotic time-series prediction tasks, achieving high accuracy, whilst enabling hardware-friendly implementation using a single VCSEL.

## Chaotic Time Series Prediction at GHz rates with a VCSEL p-SNN

Time series prediction is the task of predicting a future value in a sequence of numbers in a time-series, given the preceding values. This is a highly interesting task for different key applications, such as predictive maintenance, weather forecasting, to name but a few. However, for sequences (time-series) defined by complex or chaotic dynamics, multi-step-ahead predictions can be a very challenging task, as subsequent values are not strongly correlated to previous historic values in the sequence.

Time-series prediction has been previously investigated in traditional (non-spiking) photonic reservoir computing (RC) systems, especially in photonic implementations based upon time delay reservoirs (TDRs) using semiconductor lasers, and also including VCSELs as core nonlinear elements (see for example [32-33] and references therein). In such systems, there is an explicit optical delayed-feedback loop (typically implemented by an optical fibre link), forming recurrent connections and therefore allowing data injected in previous steps to continue circulating in the system; hence creating internal memory and influencing the output and subsequent predictions. This memory effect allows photonic TDRs to combine information from several data points in the input time series to make accurate future-value predictions [40-41].

However, to date VCSEL-based p-SNNs [34-35] do not include in its structure any intrinsic mechanism to create recurrent connections, and within them the information flows in a feed-forward manner. This circumstance means that the internal system memory inherent to photonic TDRs is not present in the VCSEL p-SNN, disabling their use for time-series prediction functionalities. We tackle this key challenge proposing an alternative data encoding approach allowing to create memory effects in the p-SNN. This is realised during the input data stage, by adding to the original data input a delayed copy of the time series. In this situation, at a given time-instant the input information contains two data input points, namely the present and a delayed value [36], using which the system is able to make multi-step-ahead predictions of a time-series. With this simple, yet very powerful approach, information about the long-range dynamics of the time-series to be predicted is made available to the p-SNN, thus effectively creating memory in the system. Notably, this is done without the need to add external fixed optical delay feedback lines, which increase complexity, require precise temporal control and can limit the total processing rate of the system. Importantly, the lack of external optical feedback lines in the p-SNN, also permits to drive the system in a so-called Extreme Learning Machine (ELM) configuration, with the information flowing in a feed-forward manner [34-35]. This readily allows the tuning of the total number of nodes in the p-SNN, without influencing the dynamical characteristics of the nodes, simply by controlling the temporal spread of the input information [34-35]. This permits the user to easily set diverse networks with higher or lower node counts depending on the complexity of the task to be processed by the p-SNN.

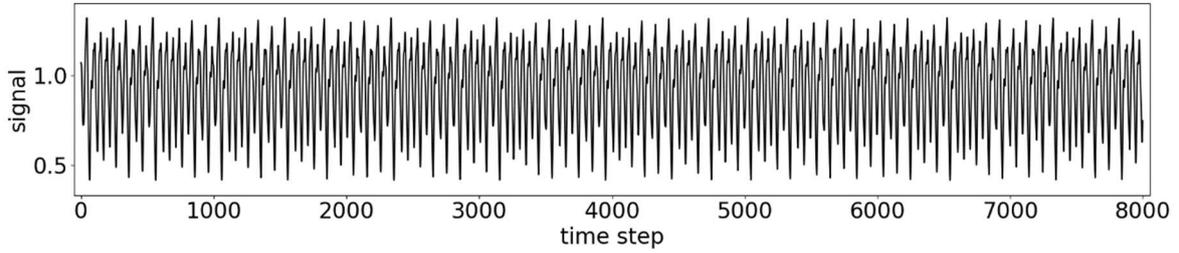

Figure 2: Time-trace providing the 8000 data points of the chaotic Mackey-Glass time-series used for the experimental investigation of the next-step ahead prediction task in the VCSEL-based p-SNN.

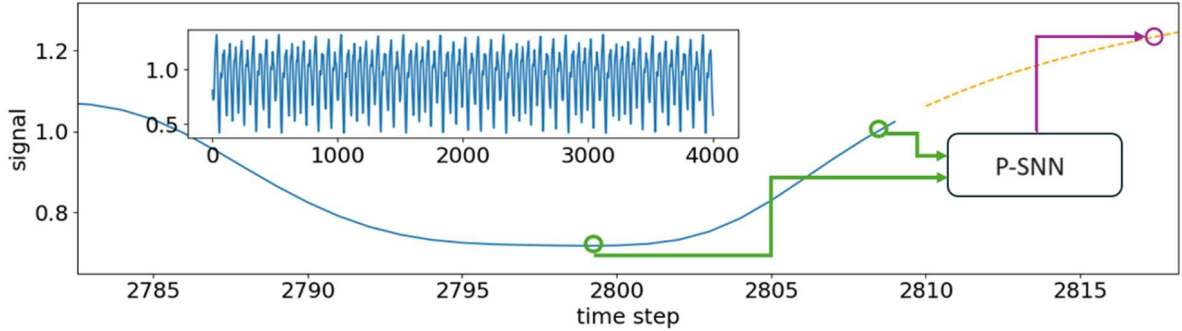

Figure 3: Operation of the p-SNN in the time series prediction task. The time series defined by the Mackey-Glass equations is shown in blue and dashed orange. The task is to use the values before a particular time step (in this case time-step 2810) to predict the value several steps ahead. To do this, the value at the time step and the value 10 steps preceding (both marked with green circles) are combined and processed in the p-SNN to find a prediction (represented with the purple circle). The inset provides the half of the full time-series.

To investigate the operation of the VCSEL p-SNN for time-series prediction, we focus on demonstrating the system's ability to perform multi-step-ahead predictions of the Mackey Glass chaotic time series, defined by the delay differential equation in equation 1 [43].

$$\frac{dy}{dt} = \frac{\beta_0 y(t-\tau)}{1 + y(t-\tau)^n} - \gamma y(t),$$

(1)

For the experimental demonstrations of this work we used a data set of 8000 inputs and targets to demonstrate the time-series prediction functionality of the VCSEL p-SNN. Figure 2 shows the 8000 time-series point sequence used as data input. To encode the inputs of the Mackey-Glass time series for their optical injection into the VCSEL p-SNN, each given data point was combined into a two-component vector with the data point delayed by 10 steps (as shown graphically in Figure 3). These resulting vectors were then multiplied by an input weight 'mask' to give the input values for each of the virtual nodes forming the p-SNN. The input mask was thus a matrix with a dimension of 2 × $N_v$, where $N_v$ describes the total number of nodes in the network, with entries drawn uniformly at random from the interval [0,1). The resulting masked input was then normalised so that all values were in the range [0,1) and were concatenated to generate the RF signal with the input data to be encoded optically (with the Mach-Zehnder Modulator in the setup) for injection into the 1300nm-VCSEL at the core of the p-SNN. The input delay of 10 steps was chosen to enable multi-step-ahead predictions resulting in predictions horizons of about 10-time units of the Mackey-Glass system [36,42].

Figs. 4(a) and 4(b) show exemplar input and output optical signals from the VCSEL p-SNN after their temporal analysis with the high-speed photodetector and real time oscilloscope in the setup. In particular, fig. 4(a) depicts the temporal varying nature of the optical input signal entering the p-SNN with the encoded Mackey-Glass data points. Fig. 4(b) shows in turn the high-speed optical spike firing signal obtained at the output of the p-SNN in response to the input shown in fig. 4(a). For the results in Figs. 4(a) and 4(b), the p-SNN was set with a node time of 250 ps and a total network node count of $N_v$ = 1024; hence yielding a total processing time of 256 ns per data-point. Figs. 4(a) and 4(b) therefore depict the input and output signals of the p-SNN when two consecutive data input points of the Mackey-Glass time series are injected into the system.

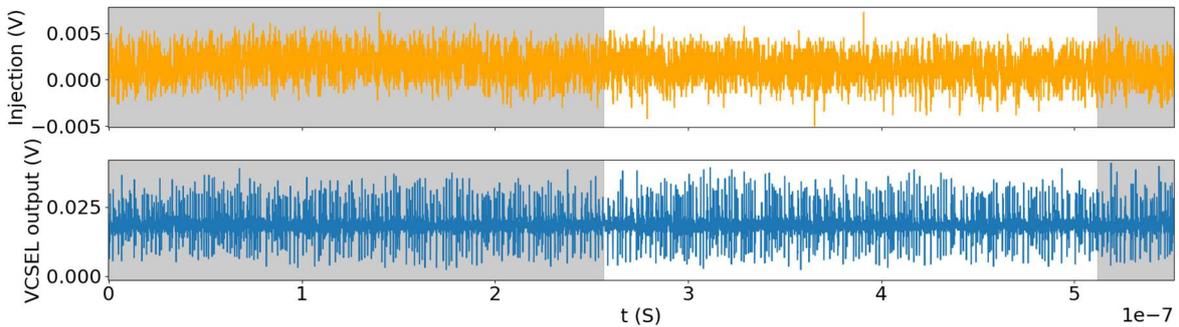

Figure 4: Photodetector reading of the intensity of light injected (top, orange) and resulting intensity output (bottom, blue) from the VCSEL p-SNN showing the triggered ultrafast (sub-nanosecond long) optical spikes. The p-SNN was set here with a total node count of $N_v$ = 1024; thus, giving a total processing time per point of 256 ns. The plots show the trace for two data points of the Mackey-Glass time-series (top, orange), lasting from 0 ns to 256 ns (shaded), and from 256 ns to 512 ns (unshaded) and the achieved optical spiking outputs (bottom, blue).

## Experimental Results

The Mackey-Glass chaotic time series prediction task was run on the VCSEL p-SNN using three n total node counts, namely $N_v$ = 256, 512 and 1024, which for a set node time of 250 ps gave processing times of 64, 128 and 256 ns per data point respectively. In all cases of analysis, the 1300nm-VCSEL at the core of the p-SNN was biased with a driving current of 3.51 mA at a temperature of 293 K [35], and the detuning of the injection frequency from the peak frequency of the subsidiary (orthogonally-polarised) mode of the VCSEL was −3.6 GHz. The average injection power in each case was 127 µW, 123 µW and 131 µW for the p-SNN architecture set with 256, 512 and 1024 nodes, respectively.

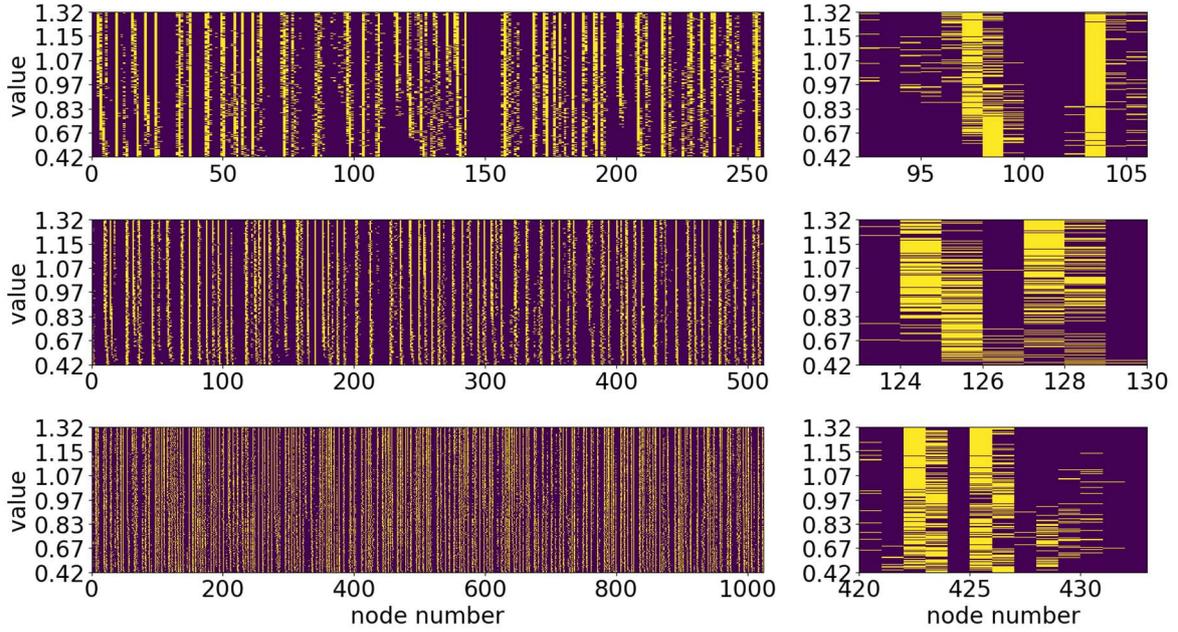

Figure 5: Temporal maps depicting the optical spike patterns delivered by the VCSEL p-SNN in response to the input data values of the Mackey-Glass time-series. Three different network architectures with total node count of $N_v$ = 256 (a, top row), $N_v$ = 512 (b, middle row) and $N_v$ = 1024 (c, bottom row) are investigated. The insets at the right-hand side provide zoomed in regions of the temporal maps that demonstrate differences in the optical spiking patterns achieved in different network nodes for different values of the Mackey-Glass time-series.

The optical spiking output signals from the p-SNN (see the exemplar in fig. 4(b)) are converted after read-out with the real-time oscilloscope in the setup, to a binary vector of length $N_v$. In this binary vector, the $i^{th}$ component is equal to '1' or '0' depending on whether the $i^{th}$ (virtual) node contained an optical spike or not (respectively). The left temporal maps in figs. 5(a), 5(b) and 5(c) merge in a single plot the binary vectors created from the optical spike patterns measured at the p-SNN's output in response to the injection of the 8000 data points of the Mackey-Glass time series. Figs. 5(a), 5(b) and 5(c) plot the obtained spiking patterns for the different p-SNNs investigated, with node counts of $N_v$ = 256, 512 and 1024, respectively. In the plots in Fig. 5 each horizontal line represents the optical output from the p-SNN in response to one data point, where a yellow (purple) pixel represents a node that contained (did not contain) an optical spike. The optical spiking patterns, that will be used to obtain the predicted values of the chaotic time-series, are sorted by the value that the p-SNN is trained to predict, given that pattern. The spiking patterns produced by by p-SNN depend on the two components of the input (the current time-series value and value delayed by 10 steps). For the p-SNN to be able to perform multi-step-ahead predictions, it must be able to uniquely map the input pairs to the value of the time-series at the targeted steps-ahead.

To train the p-SNN to perform the chaotic Mackey-Glass time-series prediction task, the following procedure was applied: first the optical spike pattern (binary) vectors obtained from the p-SNN for all input data points used for training are collected into a state matrix. Then, the state matrix and vector of target time-series values are used to find the output layer weight matrix (a vector of length $N_v$) by means of least-squares regression. After the training process, during the testing phase, the binary vectors with the p-SNN's output spiking response obtained for subsequent input data points, are multiplied by the calculated output layer weight matrix, to obtain the predicted values of the Mackey-Glass chaotic time-series. This process permits to transform the binary (spiking/non-spiking) temporal output of the p-SNN into a real number providing the next-ahead predicted value.

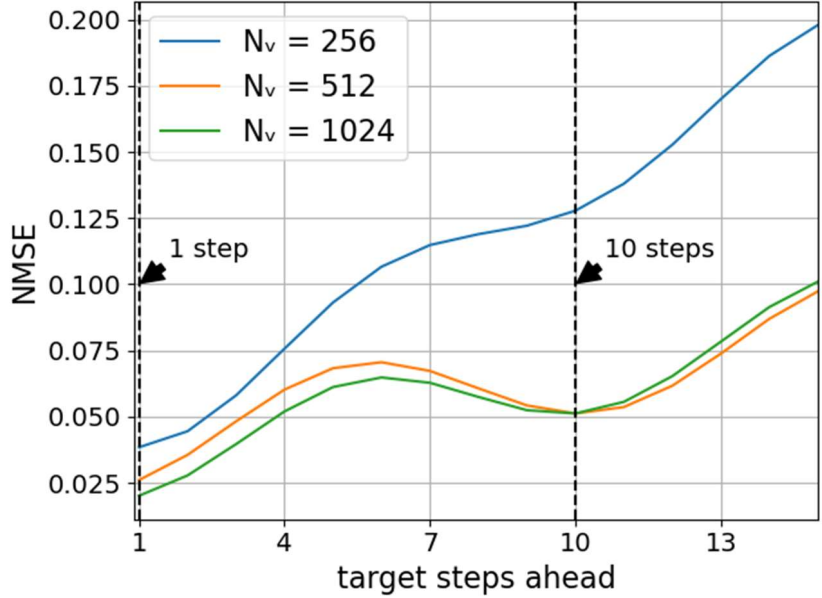

Figure 6. Multi-step-ahead prediction performance of the VCSEL p-SNN using input pairs comprised of the current value of the Mackey-Glass time-series and the value delayed by 10-steps. The x-axis is the number of steps ahead of the target and the y-axis gives the accuracy of the predictions in terms of the Normalised Mean Squared Error (NMSE). Training and testing parameters: training set size is 6000, testing set size is 1500 and the regularisation parameter is 0.03.

The performance of the p-SNN on multi-step-ahead prediction of the Mackey-Glass chaotic time-series for prediction horizons of 1 to 15 is shown in Fig. 6. The accuracy of the p-SNN has been measured using the Normalised Mean Squared Error (NMSE) which is calculated using the relation provided in equation (2). There, the true values are denoted $y$ and the prediction is denoted $\hat{y}$. Fig. 6 shows that despite the lack of intrinsic hardware memory of the p-SNN and the limited temporal information provided to it, very good prediction performance is achieved over the entire prediction range. A local minimum in the prediction error is evident around 10-steps-ahead, where a low NMSE value equal to only 0.051 is achieved. This is because the input delay of 10 used in this setup, is optimised to 10-step-ahead predictions [36]. The lowest errors are achieved for one-step-ahead predictions (with NMSE reducing to only 0.021), as the time-series values are highly correlated over this time range [43]. We selected the NMSE metric as this has widely used to analyse accuracy in traditional photonic RC systems (including those recently reported based upon VCSELs), in chaotic time series prediction tasks [32-33]. Hence, we also use it here to allow for a better comparison between the performance of the VCSEL p-SNN of this work with other photonic approaches.

$$NMSE = \frac{\sum_i (y_i - \hat{y}_i)^2}{\sigma_y^2} \qquad (2)$$

Recent works have outlined that p-SNNs, thanks to their use of spikes to process information, offer very good performance in classification tasks, even when highly reduced training dataset sizes are used [35]. This provides an advantage when compared with traditional photonic RC implementations, which typically require larger dataset sizes during the training phase to achieve good accuracy across complex tasks. Following on these recent promising reports in p-SNNs [34-35], in this work we have analysed the effect of the training data set size in the VCSEL p-SNNs performance on the Mackey-Glass time-series prediction task. Fig. 7 shows the influence of the training data set size for 10-step-ahead predictions. Here we find very good performance for training sets as small as 1600 datapoints (20% of total data points applied to the system) using the p-SNN with a 512-node network architecture. Fig. 7 also shows that for a larger network node count of 1024 nodes, larger training sets are needed to avoid overfitting.

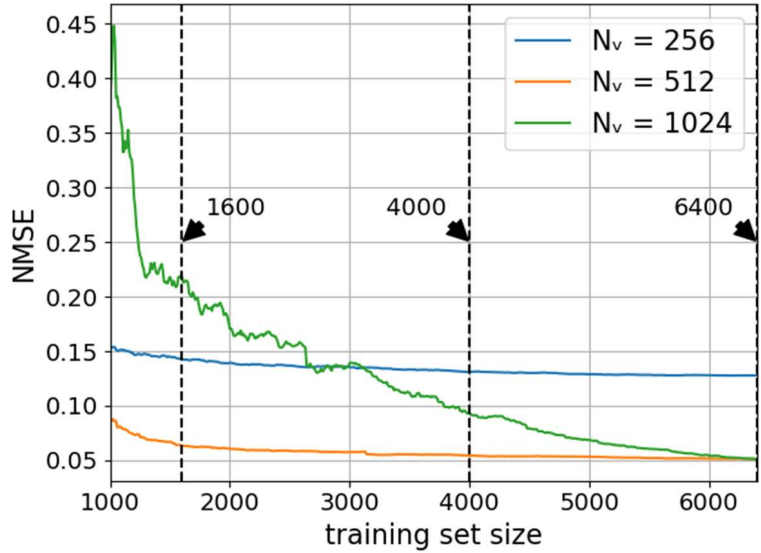

Figure 7. Performance of the VCSEL p-SNN in the 10-step-ahead prediction of the Mackey-Glass time-series as a function of the training data set size. Training and testing parameters: testing set size is 1500 and the regularisation parameter is 0.03.

To highlight the system operation, detailed results are provided in this work for three different training data sizes, namely 1600, 4000 and 6400 points (20%, 50% and 80% of total), and for the three different network architectures (with 256, 512 and 1024 nodes) investigated. Fig. 8 plots as an example of the VCSEL p-SNN performance in the Mackey-Glass time series prediction task for both 10 steps-ahead (fig. 8(a)) and 1 step-ahead (fig. 8(b)) predictions. For the results in figs. 8(a) and 8(b), a dataset formed by the first time-ordered 4000 points (50% of the total 8000 points applied) in the time-series is used for the training phase, and the last 4000 points are used during inference. For the specific cases investigated in Fig. 8 the p-SNN was configured with a total node count of 512 nodes, which at 250 ps/node, yields a total processing time of 128 ns per data point. The blue and orange traces in Figs. 8(a) and 8(b) plot respectively the original time-series and the predicted values delivered by the p-SNN. Figs. 8(a) and 8(b) shows that very good performance is obtained in both the 10 and 1 steps-ahead prediction tasks with the p-SNN. This is further illustrated in the insets in Figs. 8(a) and 8(b) showing in detail different smaller temporal sequences, highlighting the high-accuracy performance in the time-series prediction task delivered by the p-SNN, even when a very small dataset size (50% of total points) is utilised for training and with all data points fed in a timely-ordered manner. The accuracy of the p-SNN has also been measured using the correlation between true and predicted values. To that end we calculated the Pearson correlation coefficient using equation (3). There, the true values are denoted $y$ and the prediction is denoted $\hat{y}$. This measures how well the predictions depend linearly on the true value (a Pearson coefficient of 1.0 implies perfect linearity between the two variables). The plots at the right side of Figs. 8(a) and 8(b) provide the correlation between true and predicted values for the two examples considered, further highlighting the good performance of the system in predicting different step-ahead values of the Mackey-Glass time-series. A full set of experimental results on this chaotic time-series prediction task obtained for the selected training dataset sizes, 20%, 50% and 80%, and for the network architectures with 256, 512 and 1024 nodes are provided in the supporting information accompanying this article. In all cases investigated, for all cases of training data set sizes and network node count, the results highlight the successful, high-accuracy performance of the p-SNN in the chaotic time series prediction task investigated in this work.

$$\rho = \frac{cov(y, \hat{y})}{\sigma_y \sigma_{\hat{y}}} \tag{3}$$

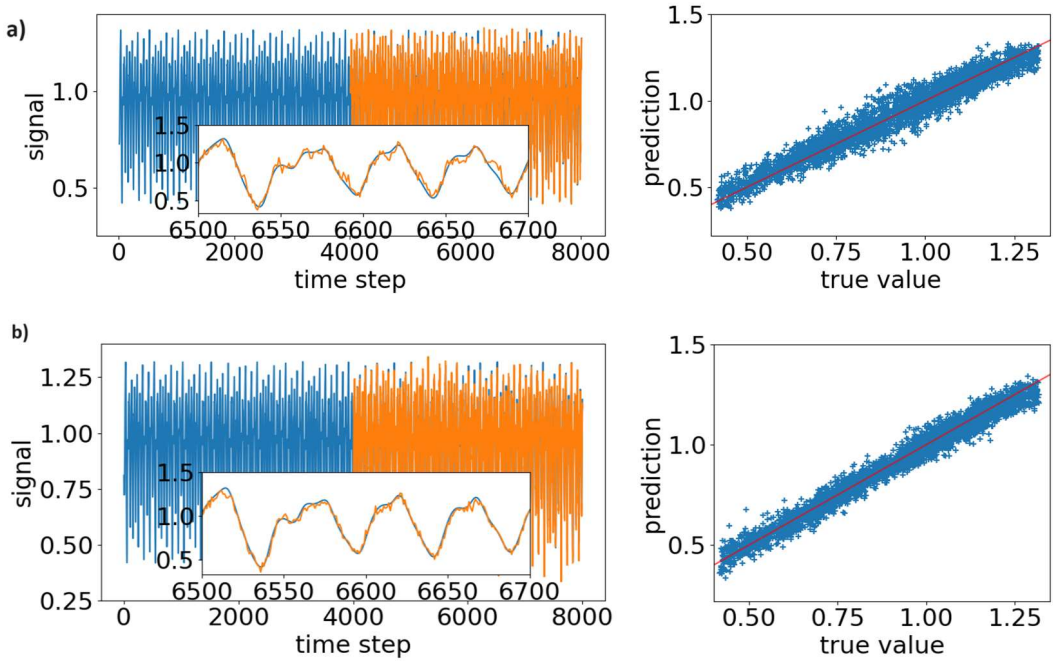

Figure 8: Performance of the p-SNN in the prediction of step-ahead values of the Mackey-Glass time series using a network with 512 virtual nodes and applying a training dataset using 4000 points (50% of total). (a) Predicting 10 steps ahead. (b) Predicting the next step ahead. Left: Predicted (orange) and true (blue) time-series values, with the first 4000 points shown in blue providing the training dataset. The inset in the left plot shows a zoomed in region between data points 6500 and 6700 of the Mackey-Glass time-series showcasing in more detail the good system performance in this prediction task. Right: Correlation between true and predicted values (in red is shown the line where prediction = truth).

Table 1: $N_v = 256$

| Training set% | nmse | correlation |
|---|---|---|
| 20 | 0.148 | 0.91 |
| 50 | 0.134 | 0.92 |
| 80 | 0.131 | 0.92 |

Table 2: $N_v = 512$

| Training set% | nmse | correlation |
|---|---|---|
| 20 | 0.066 | 0.97 |
| 50 | 0.055 | 0.97 |
| 80 | 0.051 | 0.97 |

Table 3: $N_v = 1024$

| Training set% | nmse | correlation |
|---|---|---|
| 20 | 0.213 | 0.87 |
| 50 | 0.087 | 0.95 |
| 80 | 0.051 | 0.97 |

Tables 1-3: Experimentally obtained values of NMSE and correlation coefficient for 10-steps ahead predictions for three different configured network node architectures (Nv = 256, 512, and 1024) in the p-SNN, as indicated.

Tables 1, 2 and 3 show the performance of the p-SNN in the 10 steps-ahead value prediction of the chaotic Mackey-Glass time series, measured from the calculated NMSE and correlation coefficients for the 3 specific cases of training dataset sizes of 20%, 50% and 80% (as percentage of the 8000 points used) and for network architectures with $N_v$ = 256, 512 and 1024 nodes, investigated in this work (see Supporting information for results for all 9 cases investigated in detail). The results in Tables 1-3 highlight that for all cases of analysis high-accuracy is found in the 10 steps-ahead prediction of the Mackey-Glass time-series, as per the small obtained values of NMSE (as low as NMSE = 0.051, when $N_v$ = 512) and high levels of correlation (up to 0.97 is achieved in multiple cases of analysis) between true and prediction values.

## Conclusion

This work reports experimentally for the first time the successful operation of a VCSEL-based photonic Spiking Neural Network (p-SNN) on complex time-series prediction tasks. Specifically, we demonstrate the system's accurate performance on multi-step-ahead predictions of the chaotic Mackey-Glass time series. Our results reveal that the p-SNN of this work offers excellent performance yielding high prediction accuracy whilst benefitting from ultrafast (at GHz input rates) and efficient operation (~120 µW avg. optical input power, and ~3.5mA bias current requirements) as well as from a highly hardware-friendly and inexpensive implementation using a single telecom-wavelength VCSEL and commercially-sourced telecom components. The VCSEL p-SNN of this work uses time-multiplexing to create a feed-forward network, with connectivity between neighbouring (virtual) optical spiking nodes, and the nonlinear network responses provided by the VCSEL's ultrafast neural-like excitability, its leaky-integrate and fire optical spiking dynamics, alongside the inhibitory effects induced by the refractoriness of spiking in the VCSEL. Training of the p-SNN is done only on the output layer weights, and the network operates with fixed (random) weights used in the input and hidden layers; hence reducing computational complexity while retaining highly accurate performance. Notably, our approach also eliminates the need of fixed optical delayed feedback loops to induce network recurrent connectivity and the memory needed to perform on time-series prediction tasks. Instead, a method is introduced for creating memory in the p-SNN by combining in the input data signals delayed copies of the time series; to allow past and current information to combine in the network to arrive to a successful step-ahead prediction. This is similar to typical approaches using feedforward networks or non-linear vector regression. However, here we have demonstrated a minimal approach by using only one additional time-delayed version of the input timeseries. Due to the lack of recurrent connectivity of the p-SNN the number of nodes in the network can be freely tuned as needed, without requiring any hardware modifications or controls. This permits the p-SNN to offer additional flexibility to optimise for prediction accuracy or computation speed. Moreover, the use of optical spikes to process information in the p-SNN offers inherent advantages, such as the direct use of optical spiking patterns (rather than continuous signals) to distinguish between different time-series values and accurate operation with very small training data set sizes. We demonstrate that our VCSEL-based p-SNN offers high accuracy operation (down to NMSE values as low as 0.051 for a prediction horizon of 10-steps-ahead are demonstrated experimentally) in the Mackey-Glass chaotic multi-step-ahead prediction.

## Methods

Experimental Setup:
The Photonic Spiking Neural Network (p-SNN) of this work is built using the experimental setup shown in Figure 1. At its core, the p-SNN employs a single telecom-wavelength 1300nm-VCSEL as a non-linear element for the transformation of light-encoded input data information into fast optical (neuronal) spiking signals. Standard fibre-optic telecom components are used in the setup to control, inject and analyse the system's optical signals, keeping this p-SNN approach hardware-friendly, affordable, and fully-compatible with optical communication networking and data-centre technologies. The widely employed technique of optical injection is used to introduce the input data information into the VCSEL-based p-SNN. In the optical injection arm of the setup, Continuous Wave (CW) light produced by a 1300nm tuneable laser (TL) source is intensity modulated using a 10 Gbps Mach-Zehnder (MZ) Modulator. A Power Source (PS) generates the DC bias voltage applied to the MZ Modulator, whilst a 12 GSa/s, 5 GHz-bandwidth Arbitrary Wave Generator (AWG) coupled to an RF amplifier (AMP), is used to drive the MZ Modulator with an RF signal providing the encoded input information. The optical signal in the injection arm is split by a 50:50 fibre-optic directional coupler. A first output is connected to an optical power meter to monitor the optical injection strength, whilst the second branch is split again by a 99:1 coupler. The low power (1%) branch is used to read the optical injection, and the second branch is connected to an optical circulator. The latter is used to inject the optical input signal into the 1300nm-VCSEL as well as to capture its optical output signal for subsequent analysis. The polarisation of the injected light is set using fibre polarisation controllers (PC) to ensure maximal coupling into the MZ Modulator and the VCSEL. Signal analysis was performed using a 9 GHz amplified photodetector, a 40 GSa/s, 16 GHz-bandwidth real-time Oscilloscope (OSC) and an Optical Spectrum Analyser (OSA). At room temperature (T = 293K) the threshold current of the VCSEL measured 1.4 mA, and when operated at the selected bias current of 3.51 mA, the device produced fundamental mode emission with two orthogonally-polarised peaks at 1287.65 nm and 1287.506 nm (see [35] for details on the LI curve and optical spectrum of the VCSEL). These correspond to the two orthogonal polarisations of the fundamental transverse mode of the device. Injection was made into the subsidiary mode (at 1287.668 nm) with mode-matched polarisation and a frequency detuning of -3.2 GHz. An average optical injection power of ~120 µW was incident on the VCSEL, which when run in the absence of modulation, achieved the injection locking of the device to the external signal from the TL. Fast modulation of the injection signal around these injection conditions induced fast (sub-nanosecond) optical (neural-like) spiking responses, key to the operation of the system as a p-SNN [34-35].

Mackey-Glass Timeseries:

The Mackey-Glass equations were numerically integrated using a fourth order Runge-Kutta method with Hermitian interpolation for the delayed mid-steps. The integration time step was 0.01. The timeseries was then down-sampled to time steps of 1.

Training and Testing:

The p-SNN was trained using least-squares regression. Using this method, the weights for each node were calculated using equation (4).

$$W_{out} = (S^T S + \lambda I)^{-1} S^T \boldsymbol{y} \quad , \tag{4}$$

Where the matrix $S$ is formed of the optical spiking output of the VCSEL for each data point arranged row wise, where a node containing a spike is denoted by a one, and a node without a spike by a zero. The vector $\boldsymbol{y}$ is the (column) vector of corresponding output values that the p-SNN should predict. The regularisation parameter $\lambda$ was chosen to minimize the NMSE.

Acknowledgements

The authors acknowledge this work was supported by the UKRI Turing AI Acceleration Fellowships Programme (EP/V025198/1), and by the EU Pathfinder Open project 'SpikePro'. The authors acknowledge support from the Fraunhofer Centre for Applied Photonics, FCAP.

## Author contributions

D.O.-N, A.A, and J.R. carried out the experimental analyses. J.R. and A.H. supervised the experimental activities. L.J. and K.L. contributed the memory-induced technique in the network for use in time-series prediction tasks. D.O.-N implemented the training algorithms and network memory technique. D.O-N processed the data in collaboration with A.A., J.R and J.A.J. K.L and A.H. supervised this work. All authors contributed to the writing and discussion of this manuscript.

## Competing Interests

The authors declare no competing interests.

## Materials & Correspondence

Communications about this work should be addressed to the corresponding author Dafydd Owen-Newns at dafydd.owen-newns@strath.ac.uk

## Data Availability

All data underpinning this publication are openly available from the University of Strathclyde KnowledgeBase at https://doi.org/10.15129/7350af10-809d-4c5d-b41b-0951f31e7f6c. For the purpose of Open Access, the author has applied a CC BY public copyright licence to any Author Accepted Manuscript (AAM) version arising from this submission.